\documentclass[aps,prb,twocolumn,showpacs,superscriptaddress,floatfix]{revtex4-1}
\usepackage{graphicx}
\usepackage{dcolumn}
\usepackage{bm}
\usepackage{amssymb}

\everymath{\displaystyle}

\begin{document}
\newcommand{\kp}{{\bf k$\cdot$p}\ }
\newcommand{\Pp}{{\bf P$\cdot$p}\ }
\widetext
\leftline{Version 1.0 as of \today}

\leftline{Comment to {\tt szot@ifpan.edu.pl} }

\preprint{APS/123-QED}
\title{Two-valence band electron and heat transport in monocrystalline PbTe-CdTe solid solutions with high Cd content}

\author{M. Szot}\affiliation{Institute of Physics, Polish Academy of Sciences, PL-02668 Warsaw, Poland}

\author{P. Pfeffer}\affiliation{Institute of Physics, Polish Academy of Sciences, PL-02668 Warsaw, Poland}

\author{K. Dybko}\affiliation{Institute of Physics, Polish Academy of Sciences, PL-02668 Warsaw, Poland}

\affiliation{International Research Centre MagTop, Institute of Physics, Polish Academy
	of Sciences, Aleja Lotnikow 32/46, PL-02668 Warsaw, Poland}

\author{A. Szczerbakow}\affiliation{Institute of Physics, Polish Academy of Sciences, PL-02668 Warsaw, Poland}

\author{L.~Kowalczyk} \affiliation{Institute of Physics, Polish Academy of Sciences, PL-02668 Warsaw, Poland}

 \author{P.~Dziawa} \affiliation{Institute of Physics, Polish Academy of Sciences, PL-02668 Warsaw, Poland}
 
 \author{R. Minikayev} \affiliation{Institute of Physics, Polish Academy of Sciences, PL-02668 Warsaw, Poland}

 \author{T. Zayarnyuk} \affiliation{Institute of Physics, Polish Academy of Sciences, PL-02668 Warsaw, Poland}

\author{K. Piotrowski} \affiliation{Institute of Physics, Polish Academy of Sciences, PL-02668 Warsaw, Poland}

\author{ M. U. Gutowska} \affiliation{Institute of Physics, Polish Academy of Sciences, PL-02668 Warsaw, Poland}

\author{ A. Szewczyk} \affiliation{Institute of Physics, Polish Academy of Sciences, PL-02668 Warsaw, Poland}

\author{T.~Story}\affiliation{Institute of Physics, Polish Academy of Sciences, PL-02668 Warsaw, Poland}
\affiliation{International Research Centre MagTop, Institute of Physics, Polish Academy
	of Sciences, Aleja Lotnikow 32/46, PL-02668 Warsaw, Poland}

\author{ W. Zawadzki} \affiliation{Institute of Physics, Polish Academy of Sciences, PL-02668 Warsaw, Poland}

\date{\today}
\begin{abstract}
High quality p-type PbTe-CdTe monocrystalline alloys containing up to 10 at.$\%$ of Cd are obtained by self-selecting vapor transport method. Mid infrared photoluminescence experiments are performed to follow the variation of the fundamental energy gap as a function of Cd content. The Hall mobility, thermoelectric power, thermal conductivity and thermoelectric figure of merit parameter $ZT$ are investigated experimentally and theoretically paying particular attention to the two-valence band structure of the material. It is shown that the heavy-hole band near the $\Sigma$ point of the Brillouin zone plays an important role and is responsible for the Pb$_{1-x}$Cd$_x$Te hole transport at higher Cd-content. Our data and their description can serve as the standard for Pb$_{1-x}$Cd$_x$Te single crystals with $x$ up to 0.1. It is shown, that monocrystalline Pb$_{1-x}$Cd$_x$Te samples with relatively low Cd content of about 1 at.\% and hole concentration of the order of 10$^{20}$ cm$^{-3}$ may exhibit $ZT \approx$ 1.4 at 600 K.
\end{abstract}

\maketitle

\section{INTRODUCTION}

Lead chalcogenides are IV-VI narrow gap semiconductors known for thermoelectric and mid infrared applications.\cite{pp1,pp2,pp3} An incorporation of tin into their cation sublattice leads to a decrease of the fundamental energy gap at the $L$ points of the Brillouin zone which, in the case of Pb$_{1-x}$Sn$_x$Te and Pb$_{1-x}$Sn$_x$Se  alloys, results in the band inversion giving rise to the formation of topological crystalline insulators.\cite{pp4,pp5} On the other hand, an addition of Mn, Eu, Sr or Mg ions to lead chalcogenides leads to increase of the fundamental gap and, consequently, to increase of the effective light-hole and electron masses.\cite{pp6,pp7,pp8,pp9,pp67,pp70,pp90} This enhances the thermoelectric power, lowers carrier's mobility and changes other thermodynamic characteristics.  Similar effect is expected for alloying PbTe with CdTe. In particular, it was suggested that substitution of Pb ions by Cd ones may introduce resonant level in band structure of PbTe resulting in enhancement of Seebeck coefficient through a distortion of density of states near the bottom of the conduction band\cite{pp19,pp84} correspondingly to the case of $p$-type Pb$_{1-x}$Tl$_x$Te, where the resonant level is introduced in the valence band.\cite{pp71}  However, due to different crystal structure of both materials, rock salt for PbTe and zinc blende for CdTe, these materials exhibit extremely low mutual solubility when grown from the melt.\cite{pp10,pp11,pp12,pp13} Their alloying leads usually to formation of highly symmetric (with rhombo-cubo-octahedron shape) zinc-blende precipitations of CdTe in rock-salt PbTe\cite{pp14,pp15} for sufficiently high CdTe content.\cite{pp16,pp17,pp18} The almost perfect lattice parameter matching between PbTe (6.46 ${\rm\AA}$) and CdTe (6.48 ${\rm\AA}$) results in atomically sharp PbTe/CdTe interfaces thus permitting new approach to crystal-coherent two-phase thermoelectric system. For this reason, no or only weak effect of alloying PbTe with CdTe on thermoelectric power were reported for  Pb$_{1-x}$Cd$_x$Te ($x \le 0.03$) polycrystalline samples obtained by rapid quenching or hot pressing methods. The latter growth technique is prone to the formation of CdTe nano- and microinclusions in polycrystalline matrix, which contribute to reduction of thermal conductivity and in consequence to increase of thermoelectric figure of merit parameter up to $ZT\approx1.7$ for such two-phase material.\cite{pp19,pp20,pp84}
\ \\*[1em]
\phantom{x} \hspace{1ex}In this paper we present the experimental and theoretical study of the effect of Cd ions on optical and thermoelectric properties of Pb$_{1-x}$Cd$_x$Te monocrystalline solid solutions with $x$ up to 0.1 obtained by self-selecting vapor growth (SSVG) method.\cite{pp21,pp22,pp23} A particular attention is paid to two-valence-band character of the hole transport. It is worth to note that, while polycrystals of Pb$_{1-x}$Cd$_x$Te had been available for investigations,\cite{pp63} the single crystals with the Cd content $x \ge 0.02$ were not examined. In this sense, our data and conclusions can serve as a monocrystalline reference standard for the material. This is even more important in view of the fact that Pb$_{1-x}$Cd$_x$Te has been for a long time a material of controversy, beginning with the relative positions of the light- and heavy-hole bands, the value of hole mobility, thermoelectric power and, finally, with regard to its thermoelectric figure of merit.
Precise analysis of its properties has been obscured by the two valence bands contributing to the transport properties. For this reason, PbTe and Pb$_{1-x}$Cd$_x$Te are not classical narrow gap semiconductors, although they have been considered as such.
\begin{figure*}[tph]
	\centering
	\includegraphics[width=0.85\textwidth]{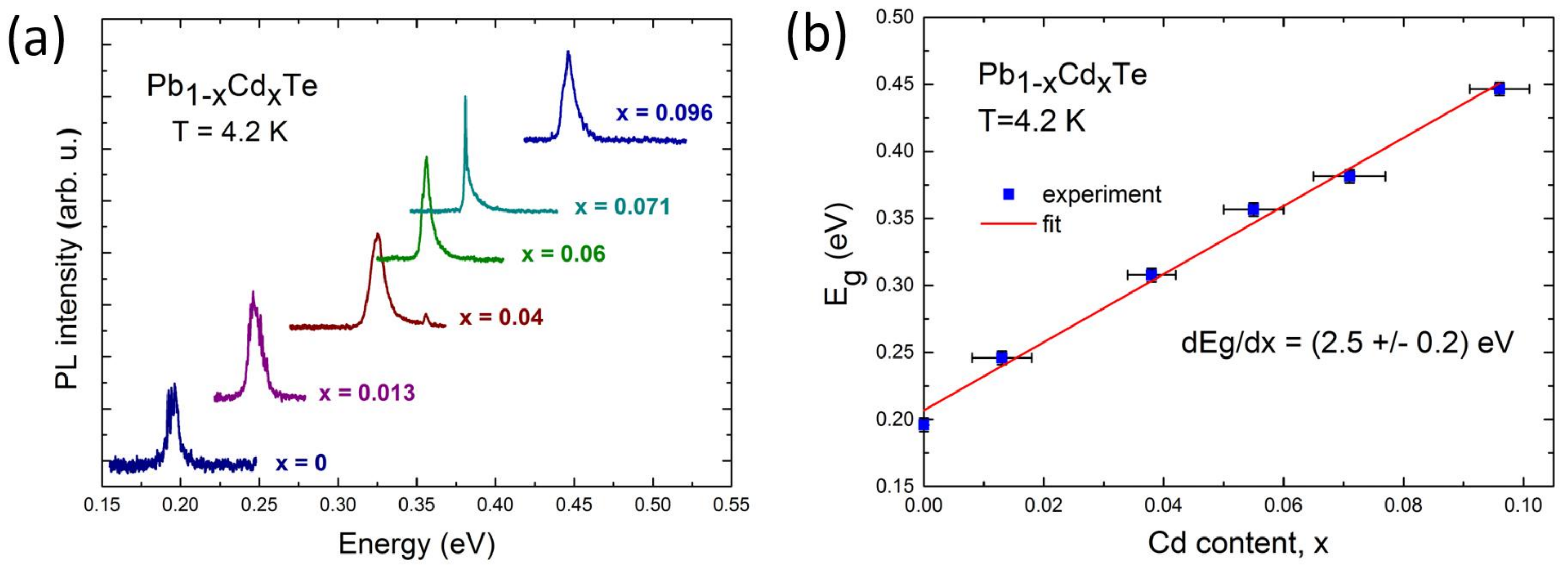}
	\caption{(a) Photoluminescence spectra of Pb$_{1-x}$Cd$_x$Te samples at 4.2 K for different Cd content $x$. (b) The dependence of energy gap E$_g$ = E$_{L_6^-}$ - E$_{L_6^+}$ of Pb$_{1-x}$Cd$_x$Te at 4.2 K on Cd content $x$.}
	\label{fig:fig1}
\end{figure*}

\section{Sample preparation and EXPERIMENT}

The SSVG method was chosen for preparation of monocrystalline Pb$_{1-x}$Cd$_x$Te as exceptionally suitable for growth of IV-VI compounds. It was confirmed in the case of Pb$_{1-x}$Sn$_x$Te and Pb$_{1-x}$Sn$_x$Se crystals.\cite{pp4,pp69,pp87} The samples were grown from polycrystalline PbTe and CdTe synthesized with excessive Te and Cd, respectively. The growth
temperatures determined using PbTe-CdTe phase diagram were about 850$\div$870 $^{o}$C, i.e. below melting point of PbTe.\cite{pp24,pp11} During the process, material placed inside the quartz ampoule evaporates from the hotter part of polycrystalline source and then condenses in its cooler region. The transport of the material was driven by small (few degrees) temperature gradient of special temperature profile inside the ampoule controlled by multi-zone electric oven, in which the growth was performed.\cite{pp22} The monocrystalline samples with Cd content up to $x$=0.102 were obtained. Monocrystals produced by SSVG are known to exhibit exceptional homogeneity. Uniform distribution of cadmium ions in individual Pb$_{1-x}$Cd$_x$Te crystals was confirmed by determination of Cd content in various parts of "as grown" crystal using scanning electron microscopy (SEM) with energy dispersive X-ray fluorescence method. The experimental uncertainty of Cd content in investigated samples does not exceed 0.6 at.$\%$.   

The Pb$_{1-x}$Cd$_x$Te samples were examined optically  by photoluminescence measurements performed in wide range of temperatures  $T$ = 4$\div$120 K using 1064 nm line of pulsed YAG: Nd laser for excitation (see Fig. 1 in main text and Figures 11, 12  in Appendix). Samples were cleaved from the host material shortly before the measurements to minimize the influence of oxidation processes on experimental results.\cite{pp25} The carrier concentration and mobility in Pb$_{1-x}$Cd$_x$Te samples were determined using measurements of the Hall effect employing standard Hall bar geometry with six contacts. To determine the Seebeck coefficient $\alpha$ (thermopower), samples were mounted between two independent heaters. Two thermocouples were used to measure the temperature gradient along the samples and to determine the Seebeck voltage. The value of thermopower for a given sample was determined as the average of measurements performed for different temperature gradients applied in both directions. The room temperature thermal conductivity $\kappa$ was measured for a few Pb$_{1-x}$Cd$_x$Te samples with different Cd content as well as for pure PbTe. The measurements were performed on separately cleaved samples with dimensions 5$\times$5$\times$5 mm suitable for measurements with Physical Property Measurment System (PPMS). Finally, the figure of merit parameter $ZT$ was also determined directly by the Harman method.\cite{pp2,pp26,pp27}

\vspace*{-1.68em}

\section{THEORY}

It is known that that the hole transport in PbTe and Pb$_{1-x}$Cd$_x$Te is governed by two valence bands of the light holes (LH) with their minima at the $L$ points and the heavy-holes (HH) near the $\Sigma$ points of the Brillouin zone. The relative energy positions of these bands are subject to a long dispute (Jaworski et al.\cite{pp60})  and  our  work contributes to the latter.  In Fig. 2 we show the band edges of the two valence bands and one conduction band of Pb$_{1-x}$Cd$_x$Te, as determined in our analysis.
The light hole (LH) valence bands in PbTe and Pb$_{1-x}$Cd$_x$Te are both nonspherical and nonparabolic. The nonsphericity is due to the  location of bands' minima at the $L$ points, while the nonparabolicity is caused by the small energy gap $E_g$ and the resulting strong {\kp} interaction with the conduction bands. Because the overall symmetry of the ellipsoid ensemble is cubic as required by rock-salt crystal symmetry, for the analysis of scattering mechanism we  approximate it by one spherical energy band with the corresponding nonparabolicity. In the {\kp} two-band model the nonparabolic dispersion is\cite{pp61}

\begin{equation}
{\cal E}=-\frac{E_g}{2}+\left[\left(\frac{E_g}{2}\right)^2+\frac{E_g\hbar^2k^2}{2m_0^*}\right]^{1/2}\;\;,
\end{equation}
where the zero of energy ${\cal E}$ is chosen at the valence band edge and $m^*_0 = 0.068m_0$ is the average  effective mass at the edge.

The anisotropy of effective mass of light holes in PbTe is characterized by $m_{||}^* / m_{\perp}^*$ = 10.8 at $T=4$ K and at $T = 300$ K. In case of the transport phenomena, for description of carriers' scattering and carriers' mobility the density of states effective mass is $m_d^*=(m_{||}^* {m_{\perp}^*}^2)^{1/3}=0.086m_0$
and the conductivity mass is $m_c^*=3/({m_{||}^*}^{-1} + 2{m_{\perp}^*}^{-1})=0.056m_0$ at $T = 300$ K. In PbTe $m_d^*$ and $m_c^*$ effective masses differ only slightly, so in our calculations we use spherical effective mass of light holes $m^*_0$ at the edge of valence band, as averaged between $m_d^*$ and $m_c^*$.

\
\begin{figure}[t]
	\centering
	\includegraphics[width=0.43\textwidth]{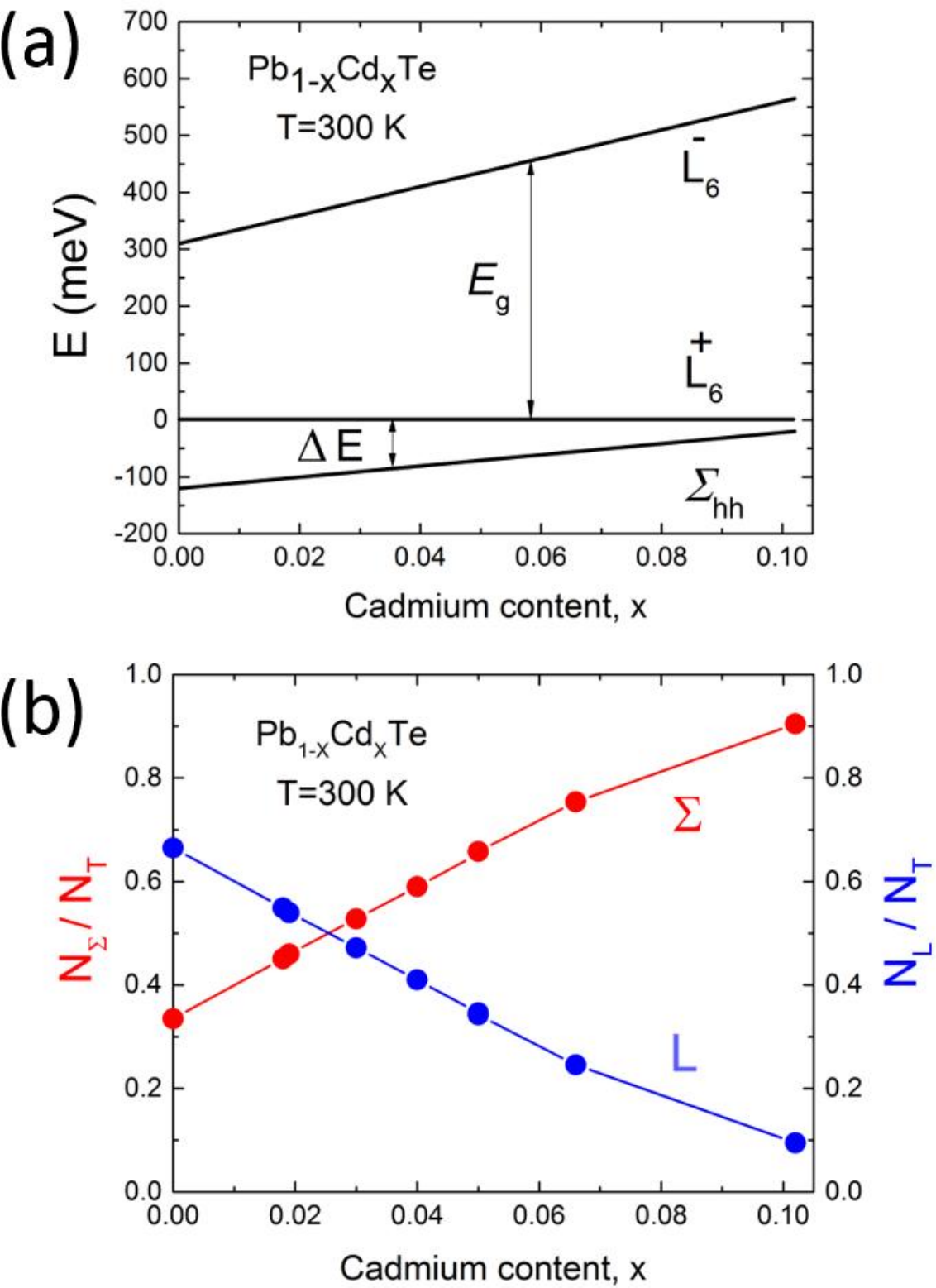}
	\caption{(a) Relative energies of two valence band edges: $\Sigma_{hh}$ and L$_6^+$ and the conduction band edge L$_6^-$ in Pb$_{1-x}$Cd$_x$Te versus Cd content $x$, as established and adjusted in the present analysis for $T = 300$ K. The light-hole effective mass is proportional to $E_g$.(b) Ratio of the heavy hole density N$_{\Sigma}$ (left scale) and light hole density N$_L$ (right scale) to the total density N$_T$ in both valence bands versus Cd content $x$, corresponding
		to the energies shown in Fig. 2(a).}
	\label{fig:fig2}
\end{figure}

The resulting energy-dependent effective mass relating velocity to pseudo-momentum $k$ is
\begin{equation}
m^*({\cal E})=m_0^*\left(1+\frac{2{\cal E}}{E_g}\right)\;\;,
\end{equation}
where
\begin{equation}
m_0^*=\frac{3E_g\hbar^2}{4P^2}\;\;,
\end{equation}
 where $P$ is the interband matrix element of momentum, (see Zawadzki\cite{pp30}). We described the composition and temperature dependence of $E_g (x,T)$ for Pb$_{1-x}$Cd$_x$Te  using our low-temperature photoluminescence data and empirical Varshni-type formula\cite{pp31} (see Fig. 12 in Appendix)
\begin{equation}
E_g (x,T)(meV)=188+2533\cdot x+\frac{0.5T^2}{T+55}\;\;
\end{equation}
However, at higher temperatures this formula slightly overestimates $E_g$, so in our calculations we use experimental value $E_g$ = 310 meV.\cite{pp66} The corresponding edge mass is
\begin{equation}
\frac{m_0^*}{m_0} =\frac{E_g (eV)}{4.08+1.6\cdot10^{-3}T}\;\;.
\end{equation}
For nonparabolic bands the transport quantities are described  in general by the integrals\cite{pp30}
\begin{equation}
<A>=\int_0^{\infty}\left(-\frac{\partial f_0({\cal E})}{\partial {\cal E}}\right)A({\cal E})k^3 d{\cal E}\;\;,	
\end{equation}
where $f_0$ is the Fermi-Dirac distribution function. In this notation the carrier density is $N = \int_0^{\infty}f_0({\cal E})\rho({\cal E})d{\cal E} = l/3\pi^2<1>$, where $l$ = $l_{lh}$ =4 for light holes. Carrier mobility is given by $\mu({\cal E}) = q\tau({\cal E})/m^*({\cal E})$, where $\tau({\cal E})$ is the total relaxation time. Various scattering mechanisms are described by separate $\tau_i$ and the total relaxation time is $\tau^{-1}=\sum_i\tau_i^{-1}$.

Carrier density is calculated using the relation $N=A_r/(qR_H )=A_r N_H$, where $R_H$ is the measured Hall coefficient, $N_H$ is the Hall carrier density and $A_r$ is the Hall scattering factor. Taking into account the mass anisotropy $K = m_{\parallel}^*/m_{\perp}^*$, there is

\begin{equation}
A_r=\frac{3K(K+2)}{(2K+1)^2}\frac{\overline{\mu^2}}{(\overline{\mu})^2}\;\;,
\end{equation}
where the average mobility $\overline{\mu}=<\mu>/<1>$. For PbTe $\frac{3K(K+2)}{(2K+1)^2}$ = 0.812 for light holes. The electric conductivity in the absence of magnetic field is
\begin{equation}
\sigma = q N \overline{\mu} = q N_H \mu_H\;\;,
\end{equation}
where $\overline{\mu} = \mu_H/A_r$ and $\mu_H$ is the Hall mobility.

The Seebeck coefficient is given by\cite{pp30}:
\begin{equation}
\alpha=\frac{k_B}{q}\left(\frac{<z \mu>}{<\mu>}-\eta\right)
\end{equation}
where $z=E/k_BT$ and $\eta$ = $E_F/k_BT$ in which $E_F$ is the Fermi energy and $<>$ denotes the integral over the band, see Eq. (5).

The thermal conductivity consists of the lattice and carrier's contributions; $\kappa=\kappa_{L}+\kappa_{c}$, where $\kappa_{L}$ at $T = 300$ K is\cite{pp64,pp40}
\begin{equation}
\kappa_{L}=\left[\frac{1}{\kappa_{Lp}}+\frac{\Omega_{0}}{4\pi\vartheta_{s}^3}x(1-x)\Gamma\right]^{-1}
\end{equation}
where $\kappa_{Lp}$ is the thermal conductivity of pure PbTe, $\Omega_{0}$ is the volume of the unit cell, $\vartheta_{s}$ is the velocity of sound, $\Gamma$ is the parameter dependent on the mass of unit cell. The carrier part of thermal conductivity is\cite{pp30}
\begin{equation}
\kappa_{c}=T\left(\frac{k_B}{q}\right)^2\sigma\left(\frac{<z^2\mu>}{<\mu>}-\frac{<z\mu>^2}{<\mu>^2}\right)\;\;.
\end{equation}

Finally, the thermoelectric figure of merit parameter is defined as
\begin{equation}
ZT=\frac{\alpha^2\sigma}{\kappa}T\;\;.
\end{equation}

The band of heavy holes is far away from other bands at the same $k$ value. In consequence it is described by the standard parabolic approximation: $ E=\hbar^2k^2/2m^*_{hh}$, where the adjusted mass is $m^*_{hh} = 0.6m_0$ (see the discussion below). The expressions for the parabolic band given above and below are formally obtained from the above description of non-parabolic ones by putting $E_g \rightarrow \infty$ (limit of non-interacting v.b. and c.b.). The number of HH ellipsoids is $l$ = $l_{hh}$ = 12. We further assume that  the HH mass does not depend on the temperature and the Cd content $x$ and take the anisotropy factor in Eq. (7) equal to unity. This is justified by a large direct gap at $\Sigma$-points. In the two band calculations the Fermi level is imposed to be the same for both bands.

For the two-band transport formulas we used the standard expressions\cite{pp62}
\begin{equation}
\mu = \frac{\overline{\mu_L} N_L+ \overline{\mu_{\Sigma}} N_{\Sigma}}{N_T}\;\;,
\end{equation}
\ \\

where $\overline{\mu_L}, \overline{\mu_{\Sigma}}, N_L, N_{\Sigma}$ are mobilities and concentrations of light and heavy holes, respectively, and $N_T= N_L + N_{\Sigma}$.
\ \\

The Seebeck coefficent is\cite{pp62}
\begin{equation}
\alpha = \frac{\alpha_L\overline{\mu_L} N_L+\alpha_{\Sigma}\overline{\mu_{\Sigma}} N_{\Sigma}}{\overline{\mu_L} N_L+\overline{\mu_{\Sigma}} N_{\Sigma}}\;\;,
\end{equation}
\ \\
where $\alpha_L$ and $\alpha_{\Sigma}$ are the corresponding quantities for the light and heavy holes, respectively.

In order to calculate  the transport quantities one needs to know and describe dominant scattering mechanisms. The relaxation times are determined by the scattering probabilities for which one needs to know the hole wave functions. Since the light hole band $L_6^+$ and the conduction band $L_6^-$ are energetically close, their wave functions are mixtures of both bands. The $L_6^+$ spin-up and spin-down bands are\cite{pp28}
$$
\Psi^{\uparrow}_v=\left[\sqrt{(1-L)}iR+\frac{L}{k}(k_z cZ-k_-dX_-)\right]\uparrow+
$$
$$
 - \frac{L}{k}\left(k_z cZ-k_z dX_+\right)\downarrow]\;\;,
$$
$$
\Psi^{\downarrow}_v=\left[\sqrt{(1-L)}iR+\frac{L}{k}(k_z cZ-k_+ dX_+)\right]\downarrow+
$$
\begin{equation}
 + \frac{L}{k}\left(k_+ cZ+k_z dX_-\right)\uparrow\;\;,
\end{equation}
where: $X_{\pm}=(X\pm iY)/\sqrt{2}$, $L={\cal E}/(2{\cal E}+E_g)$, $k_{\pm}=k_x\pm ik_y$, $R$ is the periodic amplitude of Luttinger-Kohn functions at the $L$ point of the Brillouin zone, and the normalization coefficients $c$ and $d$ fulfill the condition $c^2+d^2=1$. The effective spin-up ($\uparrow$) and spin-down ($\downarrow$) functions are indicated in the overscripts.

We consider first the polar scattering  caused by the polar interaction between longitudinal optic phonons and holes. It is described by the following formula taking into account both the screening by the hole gas as well as the phonon frequency dependence on the hole density\cite{pp28}
$$
({\tau}^{po})^{-1}(k) = \frac{kT e^2}{\hbar^2}(\frac{1}{{\varepsilon}_{\infty}}-\frac{1}{{\varepsilon}_0})\frac{F_{po} (2{\cal E}+E_g)}{\sqrt{{\cal E}+E_g}}
\sqrt{\frac{2m^*_0}{E_g {\cal E}}} =
$$
\begin{equation}
= \frac{2 kT e^2}{\hbar}(\frac{1}{{\varepsilon}_{\infty}}-\frac{1}{{\varepsilon}_0})\frac{d k}{d \cal E}F_{po}\;,
\end{equation}
\ \\
where
\ \\
\begin{equation}
F_{po} = [1-\frac{ln(1+\rho_{\infty})}{\rho_{\infty}}]
-2H[1-\frac{2}{\rho_{\infty}}+2\frac{ln(1+\rho_{\infty})}{{\rho_{\infty}}^2}]\;\;,
\end{equation}
\ \\
$H = L(1-L)$ and $\rho_{\infty} = 4k^2{\lambda_{\infty}}^2$ where
\begin{equation}
\frac{1}{{\lambda_{\infty}}^2} = \frac{2le^2}{\pi \varepsilon_{\infty}}\langle\frac{(E_g+2{\cal E})}{(E_g+{\cal E}){\cal E}}\rangle
\end{equation}
\ \\
and $\lambda_{\infty}$ is the screening length for the high-frequency dielectric constant. We use for polar optical scattering the relaxation time approximation because $\hbar\omega_{op}/k_BT \simeq 1/2$. For lower temperatures (nonelastic scattering) one needs variational methods.\cite{pp42}
\ \\

The other three important scattering mechanisms are caused by the deformation potentials with acoustic and optic phonons and by alloy disorder. For acoustic phonons the relaxation time is\cite{pp28}
\ \\
\begin{equation}
({\tau}_{ac})^{-1}(k) = C_{ac}(E^v_{ac})^2F_{ac}\;,
\end{equation}
\
\begin{equation}
C_{ac}=\frac{k_BT}{\pi \hbar c_1}\frac{\partial k}{\partial{\cal E}} k^2\;,
\end{equation}
where the acoustic deformation potentials are $E^v_{ac}=E^c_{ac}$ and $c_1$ is the combination od elastic moduli related to the averaged velocity of the sound wave.\cite{pp28,pp65}
\ \\

For nonpolar optical scattering there is\cite{pp43}

\

\begin{equation}
({\tau}_{npo})^{-1}(k) = C_{npo}(E^v_{npo})^2F_{npo} \;,
\end{equation}
\
\begin{equation}
C_{npo}=\frac{\pi kT \hbar}{\varrho (\hbar \omega_{op})^2 a^2_0}\frac{\partial k}{\partial{\cal E}} k^2 \;,
\end{equation}
in which $E^v_{npo}=E^c_{npo}$ are the nonpolar deformation potentials, $\hbar \omega_{op}$ is the energy of the optic phonon, $\varrho$ is the crystal density, $a_0$ is the lattice constant.

\begin{figure}[t]
	\centering
	\includegraphics[width=0.42\textwidth]{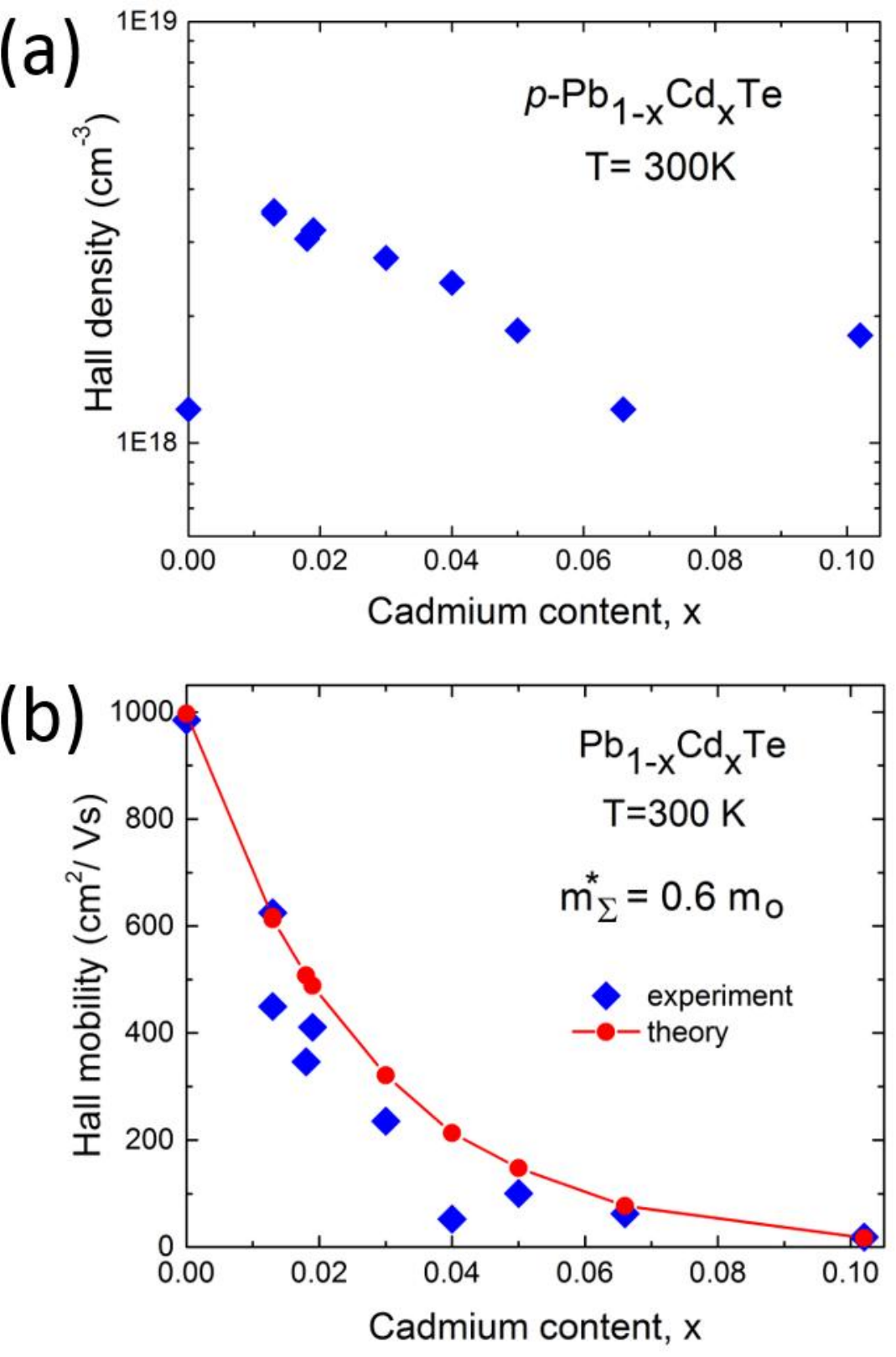}
	\caption{Carrier density (a)  and Hall mobility (b)  in monocrystalline Pb$_{1-x}$Cd$_x$Te alloy at room temperature averaged over light and heavy hole bands versus Cd content $0$ $\leq$ $x$ $\leq$ $0.102$. Full diamonds are experimental, circles are theoretical, the line is a guide to the eye.}
	\label{fig:fig3}
\end{figure}

Finally, the alloy disorder mechanism is described by\cite{pp32,pp33}
\begin{equation}
({\tau}_{ad})^{-1}(k) = C_{ad}(E^v_{ad})^2F_{ad} \;,
\end{equation}
\
\begin{equation}
C_{ad}=\frac{4x(1-x)}{\pi \hbar \Omega}\frac{\partial k}{\partial{\cal E}} k^2 \;.
\end{equation}
where $E^v_{ad}$ and $E^c_{ad}$ are the matrix elements for the valence and conduction bands.
The above three modes contain the common factor $F_i$
\begin{equation}
F_{i} = [1-L(1-y)]^2-yH\frac{8}{3}\;\;,
\end{equation}
where $y=E^v_i/E^c_i$.

For the parabolic band of heavy holes there is  H=0 in Eq. (17) and $F_i$=1 in Eq. (25). According to the Wiedemann-Franz law, see Eq. (10), the carrier  thermal conductivity is proportional to the electrical conductivity. Since the latter is proportional to $\mu_{H}$ and for the HH band this mobility is quite low, the HH contribution to the thermal conductivity is negligible.

\vspace*{-1em}

\section{RESULTS AND DISCUSSION}

Our X--ray diffraction measurements revealed that the Pb$ _{1-x}$Cd$_x$Te samples grown by SSVG method maintain rock salt structure and high crystal quality with the single phase (see Figs. 9 and 10 in Appendix). The lattice parameter of Pb$_{1-x}$Cd$_x$Te monocrystals decreases with $x$ following the Vegard's law with coefficient
 $da_0/dx$ = -0.43 ${\rm \AA}$.\cite{pp21} The value of this coefficient is 
higher than those obtained earlier by Rosenberg et al.\cite{pp10} and Leute et al.\cite{pp11}(-0.30 ${\rm\AA}$, and -0.40 ${\rm\AA}$, 
respectively) which indicates that the number of Cd ions located in 
the interstitial positions in crystals grown by SSVG method is 
considerably lower then those obtained by other 
methods.\cite{pp13,pp18,pp20}
Results of low temperature photoluminescence measurements in mid-infrared region shown in Fig. 1 indicate that substitution of Pb ions by Cd ones causes linear increase of the fundamental energy gap with $dE_g/dx$ = 2.5 eV in investigated range of $x$. Thus, over twofold increase of $L$-point energy gap is observed for $x\approx 0.1$. This behavior is related to the considerably higher energy gap of CdTe (1.6 eV at 4 K) compared to PbTe. Both effects, the decrease of lattice parameter and increase of the band gap of PbTe-CdTe mixed crystals are in agreement with theoretical predictions based on density functional theory and tight-binding approach.\cite{pp6} The energy gap of Pb$_{1-x}$Cd$_x$Te increases almost linearly  with temperature: $dE_g/dT \approx$ 0.4$\div$0.5 meV/K, which is similar to that observed for PbTe.\cite{pp34}
\begin{figure}[t]
	\centering
	\includegraphics[width=0.42\textwidth]{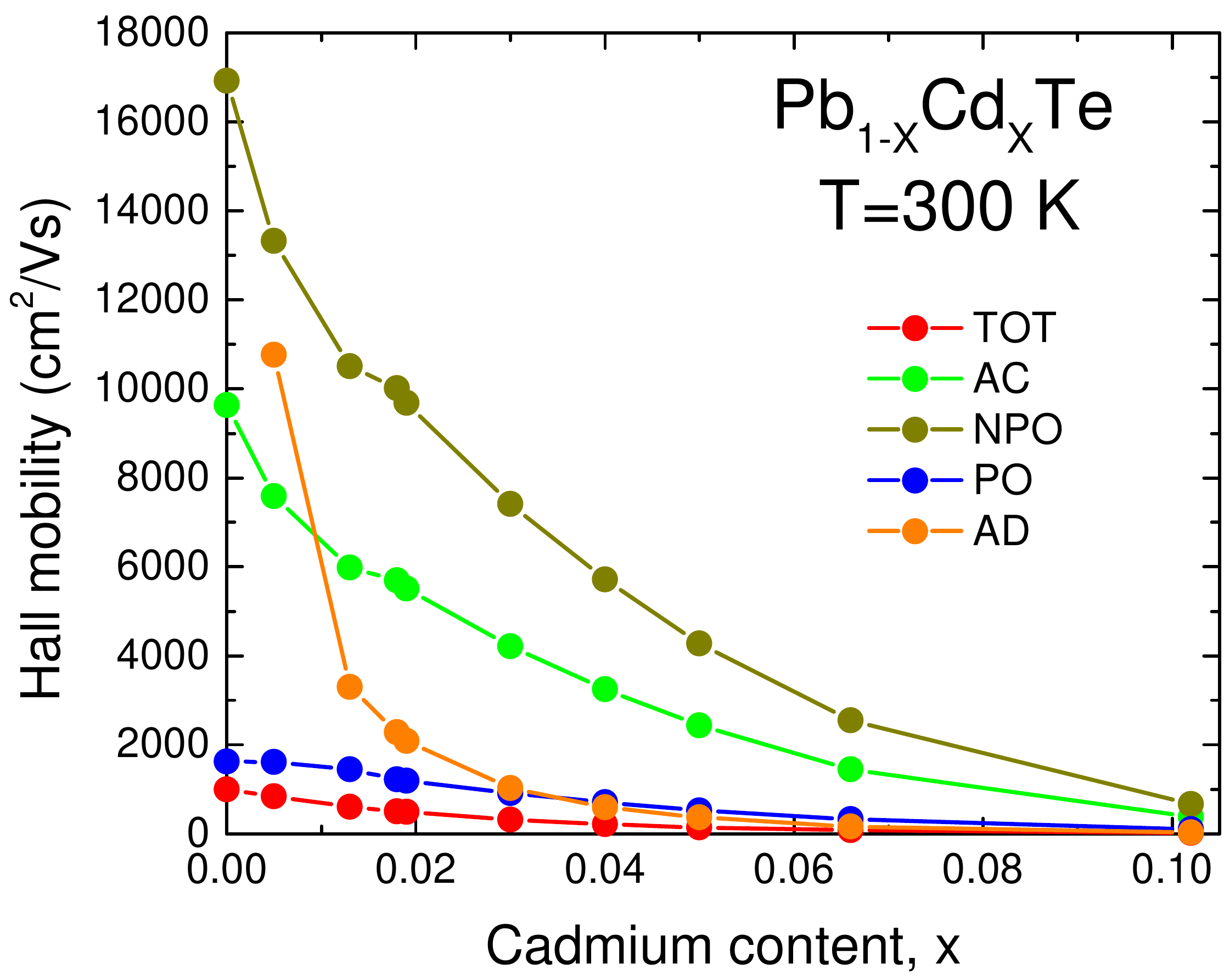}
	\caption{\label{fig:epsart}{Calculated contributions of individual scattering modes to the total hole mobility in Pb$_{1-x}$Cd$_x$Te at $T = 300$ K versus Cd content $x$. TOT- total mobility, PO-polar optical interaction, AC- acoustic deformation potential, NPO - nonpolar optical interaction, AD - alloy disorder scattering. Calculations performed using the hole density obtained experimentally for given $x$ - see Fig. 3(a). }}
	\label{fig:fig8}
\end{figure}

Our results report mostly transport properties of monocrystalline $p$-type Pb$_{1-x}$Cd$_x$Te at room temperature with the Cd content $0 \le x \le 0.102$. The first step in the theoretical description of the data is to separate contributions from the light and heavy hole valence bands. This is intimately related to the energy separation of the band edges and $m^*_{hh}$ value. We first take a tentative separation of these bands for PbTe and heavy holes mass and tentatively divide the measured value of the total hole density into the LH and HH contributions. Next we calculate the Fermi energies E$_F^{lh}$ and E$_F^{hh}$ using one the formulas for LH band and its simplified version for the HH band with $\Delta E$ included. Then, by the iteration procedure we change the density distribution between the two bands and repeat the procedure until E$_F^{lh}$ and E$_F^{hh}$ become equal. This completes the distribution of holes between the two valence bands for a given value of $x$ and the total hole density for assumed $\Delta E$ and value of $m^*_{hh}$.\cite{pp85,pp86} Next all the considered transport effects are  calculated adjusting the other transport parameters until we reach overall optimal agreement between the experiment and theory. This procedure has the merit of keeping the theory close to the experimental reality.
The specific band structure is illustrated in Fig. 2(a), which shows the band ordering of valence bands in PbTe and Pb$_{1-x}$Cd$_x$Te near the $L$ and $\Sigma$ points of the Brillouin zone.
\
\begin{figure}[t]
	\includegraphics[width=0.43\textwidth]{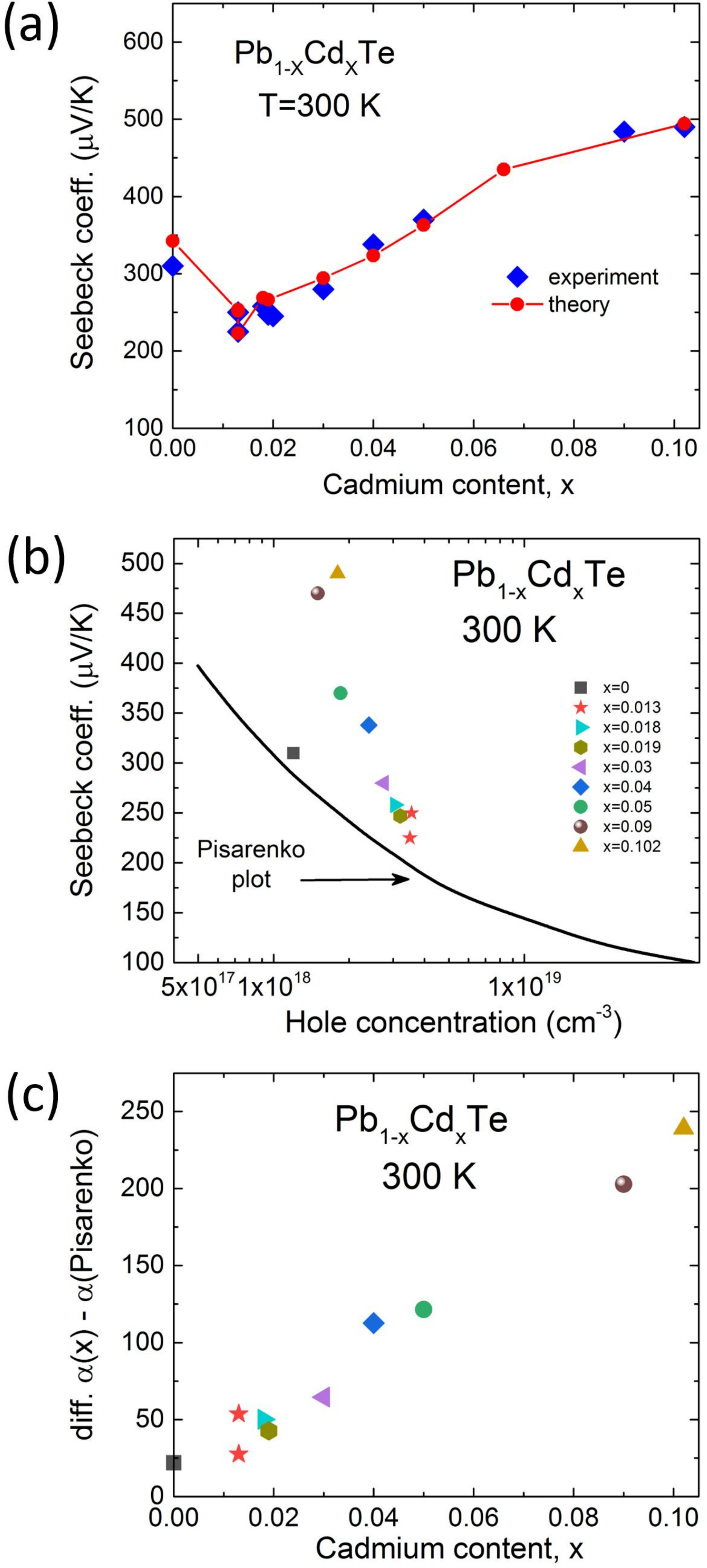}
	\caption{\label{fig:epsart}{The Seebeck coefficient of Pb$_{1-x}$Cd$_x$Te at room temperature versus Cd content (a), carrier concentration (b). Gain of Seebeck coefficient related to the reference Pisarenko plot (c). Each experimental point is plotted for a sample with known $x$ value and measured hole density - see Fig. 3(a). Line joins the theoretical points calculated for the corresponding parameters.}}
	\label{fig:fig4}
\end{figure}

  Electrical characterization reveled p-type conductivity of Pb$_{1-x}$Cd$_x$Te for all cadmium contents studied with room temperature hole densities varying between (1.2$\div$3.6)$\times$10$^{18}$ cm$^{-3}$. The  hole density for $x$ = 0.02 sample is about 3 times higher
compared to pure PbTe.  However, further increase of Cd content results in lowering the hole concentration from 3.6$\times$10$^{18}$ cm$^{-3}$ to 1.8$\times$10$^{18}$ cm$^{-3}$
 for $x\approx$ 0.1 as shown in Fig. 3(a). The determined experimental Hall mobility of Pb$_{1-x}$Cd$_x$Te versus the Cd content $x$ and its theoretical description for various hole concentrations is shown in Fig. 3(b). It is seen that the mobility diminishes  quite strongly with $x$. The main reason for the decrease is that, as seen in Fig. 2(a), when the Cd content grows, the HH band approaches in energy the LH band. In consequence, the contribution of HH to the total conduction increases and N$_L$ decreases which is shown in Fig. 2b. This lowers the overall mobility (Eq. 13) since the HH mobility is much lower than that of LH due to its much heavier mass.
 The second reason for the fall of mobility is that, as again seen in Fig. 2a, the fundamental energy gap $E_g$ between the L$_6^-$ and L$_6^+$ bands grows with the content $x$. This results in the enhancement of the light-hole mass, see Eq. (5), which further lowers the LH mobility in  the L$_6^+$ band. Calculated contributions of individual scattering modes to the total hole mobility in Pb$_{1-x}$Cd$_x$Te at $T=300$ K versus Cd content $x$ are shown in Fig. 4. Using the results of our calculations we can conclude that main scattering mechanisms at room temperature are: the interaction with optical phonons - for all Cd contents studied, and alloy disorder scattering in the case of samples with $x\geq 0.01$. It is seen in Fig. 3(b) that our theoretical description of the measured mobility values is quite successful.
\ \\
\begin{figure}[t]
	\centering
	\includegraphics[width=0.42\textwidth]{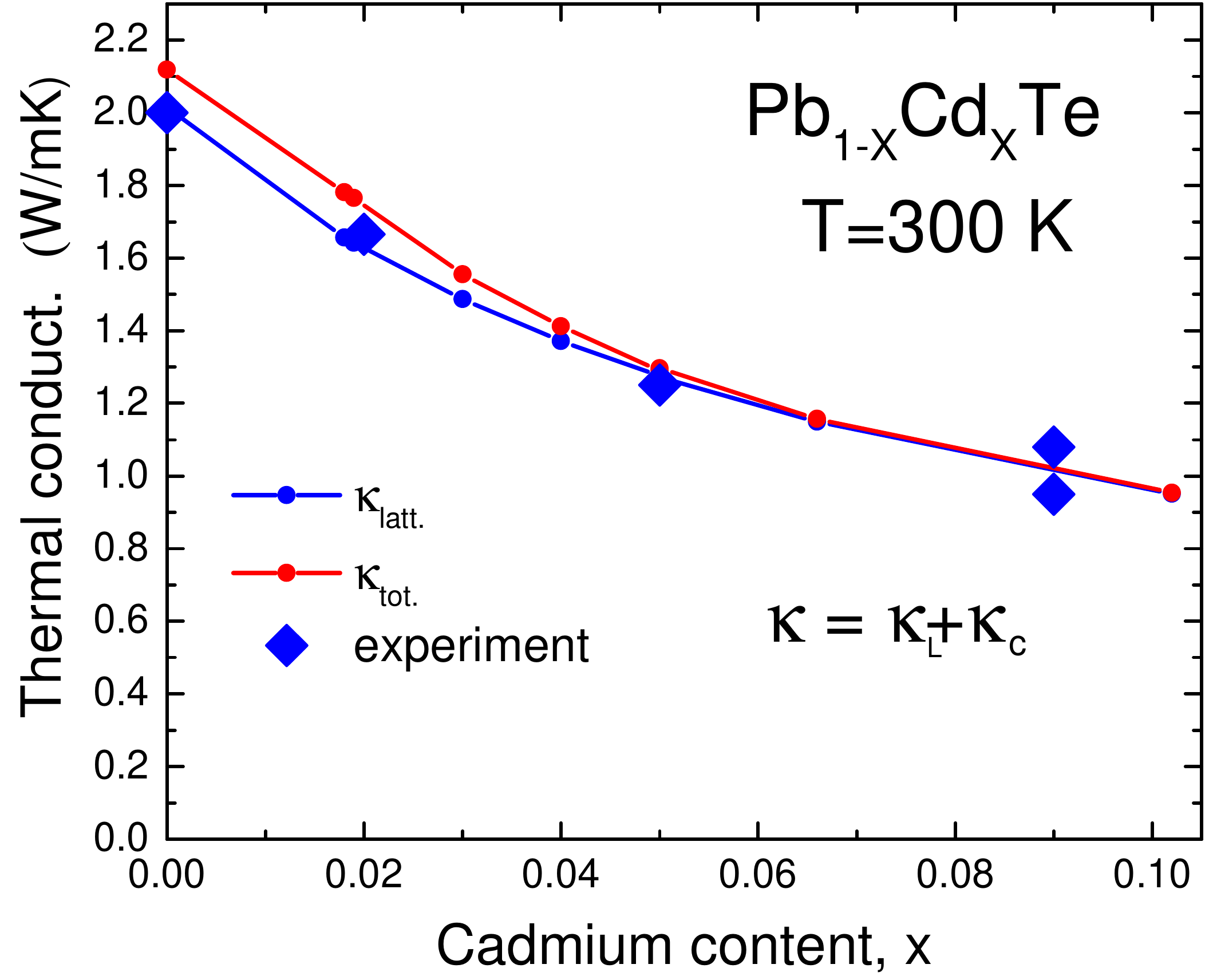}
	\caption{\label{fig:epsart}{Thermal conductivity of Pb$_{1-x}$Cd$_x$Te versus Cd content. Squares are our experiment, the lower line is theoretical for lattice contribution to $\kappa$, the upper one takes additionally into account free hole contributions to total $\kappa$.}}
	\label{fig:fig5}
\end{figure}
Figure 5(a) shows experimental and theoretical results on the Seebeck coefficient $\alpha$ of Pb$_{1-x}$Cd$_x$Te as function of the Cd content. Similarly to Fig. 3, the experimental points are somewhat scattered since they correspond to different total hole densities determined experimentally. On the whole, $\alpha$ grows both experimentally and theoretically with increasing addition of Cd reaching 490 $\mu$V/K for highest $x$. However, observed enhancement (see Fig. 5(b)) is faster if compare to pure PbTe with decreasing hole concentration (so-called Pisarenko plot\cite{pp35,pp36,pp37} - solid line in Fig. 5(b)). It is clearly seen in Fig. 5(c), where differences between measured values of Seebeck coefficient and these expected for PbTe with appropriate hole densities as a function of Cd content $x$ are shown.
It follows from the two-band formula (14) that the HH band contributes little to the total $\alpha$ because of the low HH mobility. It is difficult to judge the behavior of $\alpha$ from the exact but general expression (9), but it is possible to evaluate it more explicitly from an approximation applicable to partly degenerate carrier gas.\cite{pp30} In the linear approximation in $T$ one obtains $\alpha \sim Tm^*$/N$^{2/3}$, where $m^*$ is the density of states effective mass at the Fermi level defined above in Eq. (2). It then follows that the thermoelectric power grows with the increasing mass and diminishing LH density. This is roughly what one observes in Fig. 5. The calculation with the use of exact formula (9) gives the  correct description of the data. 
\begin{figure}[t]
	\centering
	\includegraphics[width=0.42\textwidth]{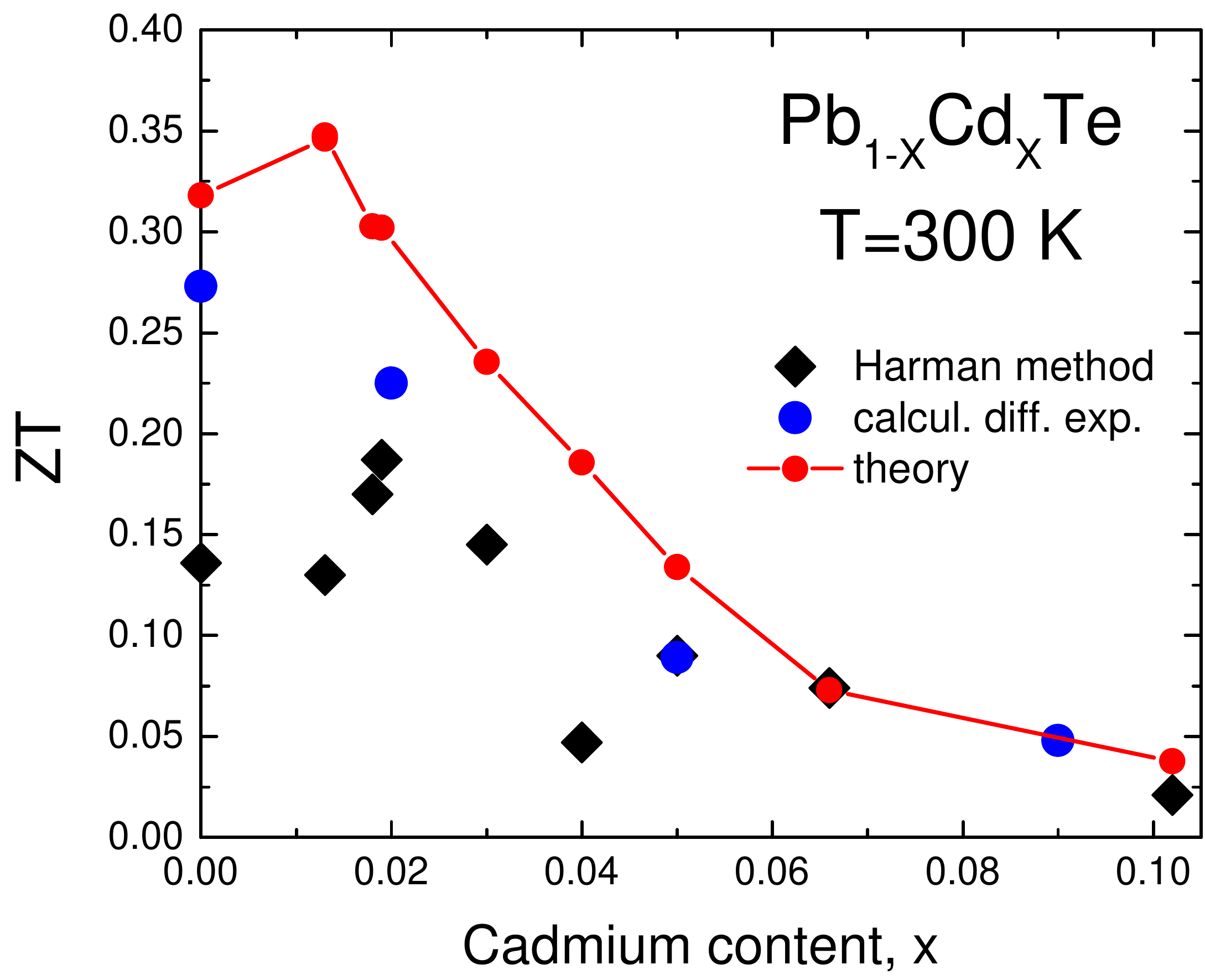}
	\caption{\label{fig:epsart}{Thermoelectric figure of merit parameter $ZT$ of monocrystalline Pb$_{1-x}$Cd$_x$Te at room temperature versus Cd content $x$ as determined experimentally by two methods, see text. The theoretical line joins points calculated for given $x$ and measured Hall carrier density.}}
	\label{fig:fig6}
\end{figure}
In Fig. 6 we show the experimental and theoretical thermal conductivity $\kappa$ of our Pb$_{1-x}$Cd$_x$Te single crystals versus $x$. It is seen that addition of Cd decreases the total $\kappa$ to 1 W/mK for samples with $x$ = 0.09. The mobile hole contribution to the total $\kappa$ is not large but not negligible either. It follows from the formula (11) expressing the Wiedemann-Franz law that the low conductivity $\sigma$ of HH band practically suppresses its contribution to $\kappa$. For this reason the contribution of $\kappa_c$ disappears at higher $x$. The dominating lattice part was calculated early using the theory of Callaway et al. and the more recent treatment of Tian et al..\cite{pp64,pp40} The general decrease of $\kappa_L$ with growing $x$ is due to the increasing alloy disorder scattering of acoustic phonons. The overall theoretical description of the data is very good.
\

Finally, Fig. 7 illustrates the measured and theoretical thermoelectric figure of merit parameter $ZT$ of Pb$_{1-x}$Cd$_x$Te versus Cd content $x$ at $T = 300$ K. The data presented in Fig. 7 were obtained for samples with different hole concentration (see Fig. 3(a)) resulting from slightly varying stoichiometry. Our experimental data are of two kinds. The full circles show values obtained by measuring separately $\sigma, \alpha$, and $\kappa$ and combining them into $ZT$ according to the formula (12). On the other hand, the diamonds indicate results of $ZT$, as obtained directly by the method of Harman. It can be seen that the method of Harman gives consistently lower values of $ZT$. As in all above figures, the theoretical line is obtained by joining the points calculating the corresponding quantity for given values of $x$ and the measured hole density N$_H$. Thus, the apparent theoretical maximum of $ZT$ at $x$ = 1.3  at.  $\%$ is due to the high hole density 3.6 $\times$$10^{18}$ cm$^{-3}$ taken for the calculation. On the whole, the agreement between the theory and experiment should be considered as quite reasonable. It is seen that the overall increase of $\alpha$ cannot compensate the strong fall of the mobility shown in Fig. 3(b) thus, the $ZT$ parameter decreases for higher Cd content. However, our calculations of $ZT$ dependence on carrier concentration (see Fig. 8) performed for $x$=0.01 sample suggest that, at $T = 300$ K, in this range of carrier's density $ZT$ parameter may change significantly. Optimal hole concentration  maximizing $ZT$ at room temperature should be $\approx$ 6$\times$10$^{18}$ cm$^{-3}$. Moreover, extending our model for higher hole densities and temperatures we found, as expected for PbTe-based thermolectrics, that $ZT$ increases rapidly reaching $ZT$ $\approx$ 1.4 at $T = 600$ K for $N_h\approx 10^{20}$ cm$^{-3}$.

\begin{figure}[t]
	\centering
	\includegraphics[width=0.42\textwidth]{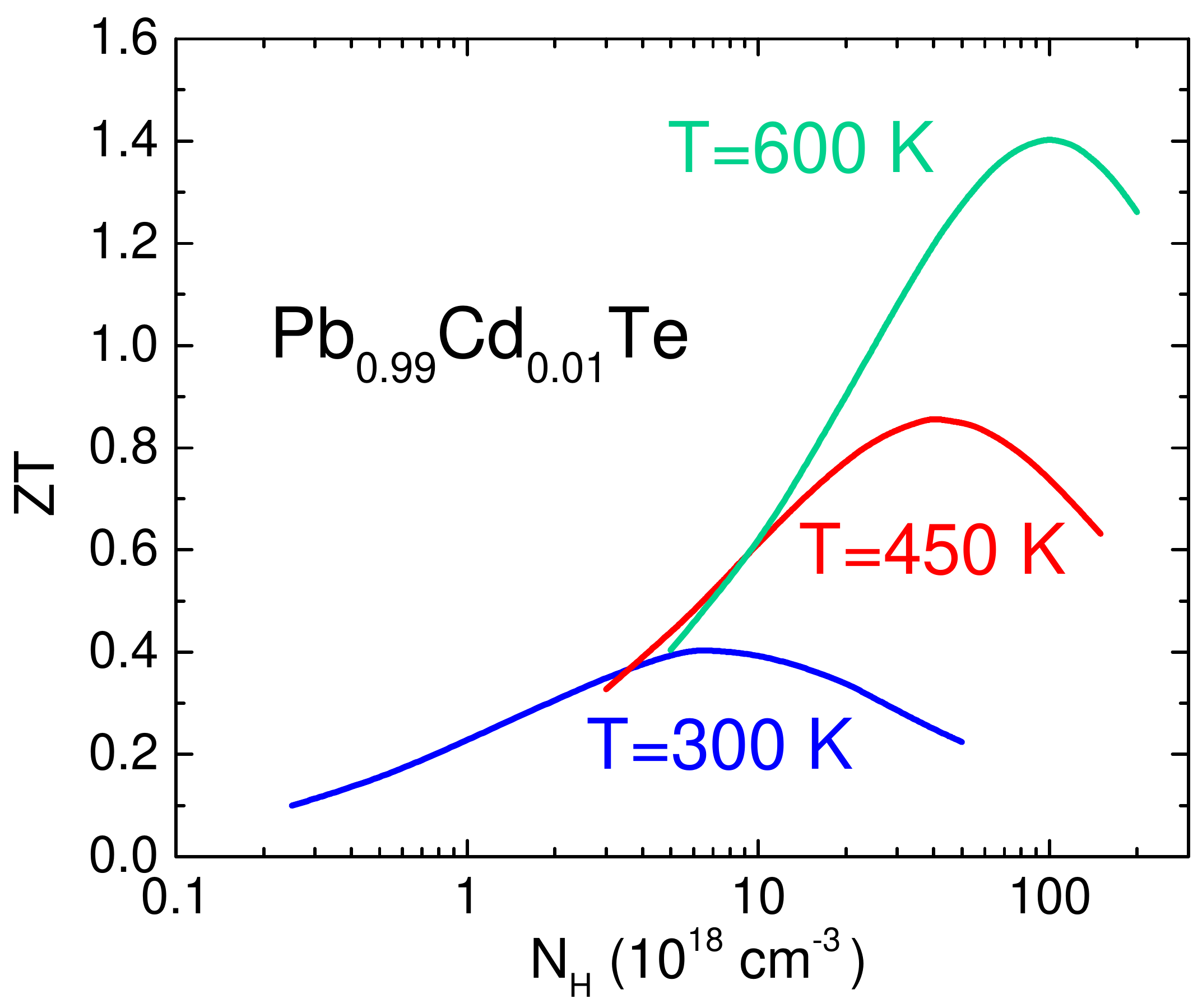}
	\caption{\label{fig:epsart}{Theoretical thermoelectric figure of merit parameter $ZT$ of Pb$_{0.99}$Cd$_{0.01}$Te versus carrier concentration at $T$ = 300, 450 and 600 K.}}
	\label{fig:fig9}
\end{figure}

\vspace*{-1em}

\section{SUMMARY}

We prepared the monocrystalline Pb$_{1-x}$Cd$_x$Te samples up to $x\approx0.1$  using SSVG  method. Our monocrystals can serve as a reference for heavily doped, polycrystalline PbTe-CdTe materials with Cd content  limited to $x\approx0.03$, which exhibit high thermoelectric figure of merit ($ZT$ = 1.2$\div$1.7).\cite{pp19,pp20,pp84}  The optical $L$-point energy gap of Pb$_{1-x}$Cd$_x$Te monocrystals determined from mid-infrared photoluminescence measurements grows with increasing Cd content and temperature with $dE_g/dx$ = 2.5 eV and $dE_g/dT \approx$ 0.4$\div$0.5 meV/K, respectively. Our results also indicate that alloying PbTe with Cd causes twofold  enhancement of the Seebeck coefficient (up to $\approx500$ $\mu$V/K), which we attribute to the increased energy gap and growing contribution of $\Sigma$-band  heavy holes. In parallel, we observe the reduction of thermal conductivity to 1 W/mK in Pb$_{1-x}$Cd$_x$Te samples with highest $x$ related to additional phonon scattering caused by substitutional Cd ions. As to the thermoelectric figure of merit parameter $ZT$, the benefit of more favorable thermopower and thermal conductivity is counteracted by strong reduction of the hole mobility for samples with higher $x$ due to increased role of heavy holes. On the other hand, our theoretical analysis indicates very strong dependence of $ZT$ on the hole concentration and temperature for  samples with relatively small $x$. Thus, optimization of carrier density by additional doping or proper post-growth annealing may significantly improve $ZT$ parameter of Pb$_{1-x}$Cd$_x$Te monocrystals. 

\vspace*{-1em}

\begin{acknowledgments}
	The research was partially supported by the National Centre for Research and Development (Poland) through Grant TERMOD No. TECHMATSTRATEG2/408569/NCBR/2019 and by the Foundation for Polish Science through the IRA Programme co-financed by EU within SG OP.

\end{acknowledgments}
\begin{figure*}[b]
	\includegraphics[width=0.45\textwidth]{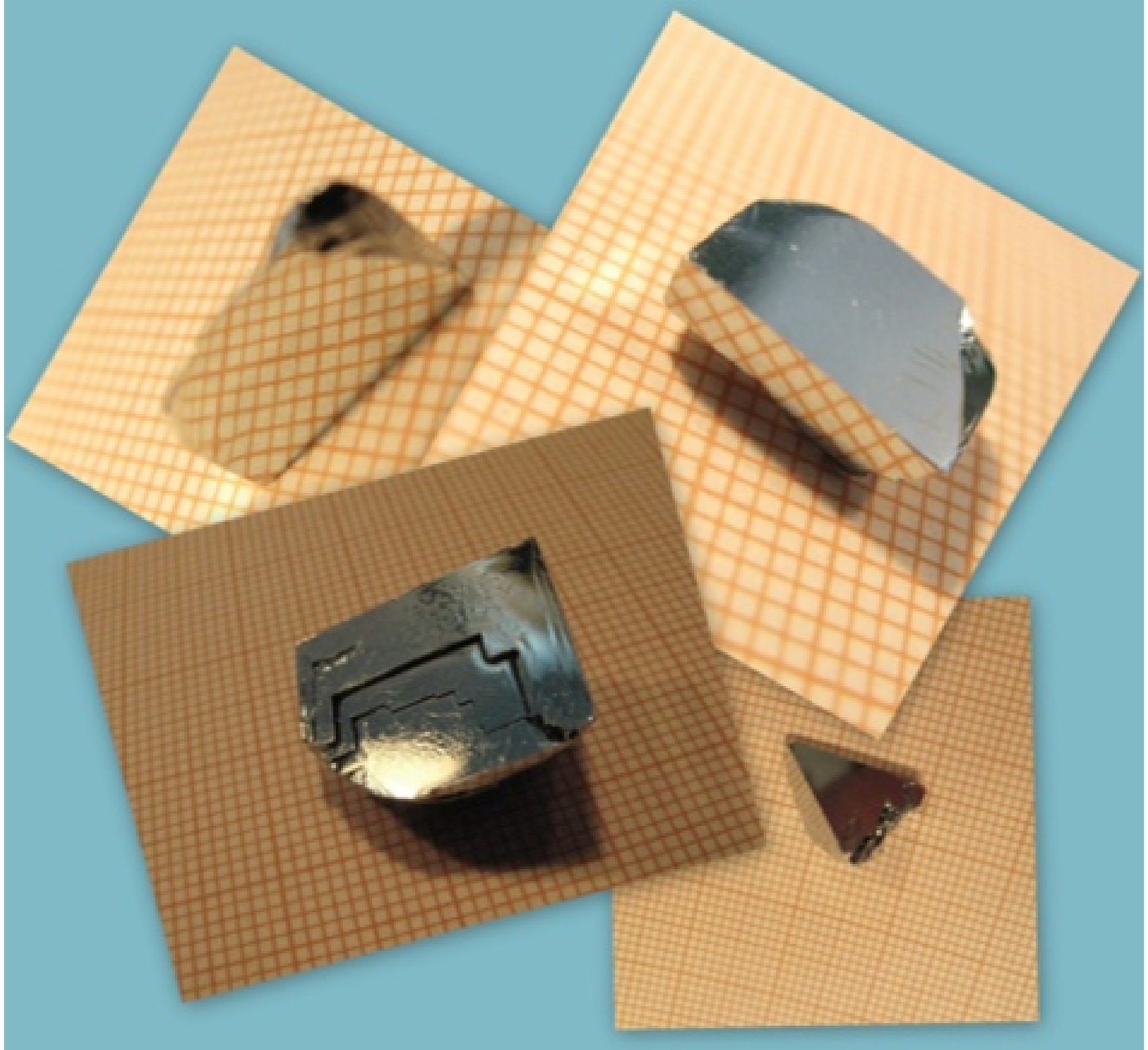}
	\caption{ Pb$_{1-x}$Cd$_x$Te monocrystals obtained by SSVG method.}
	\label{fig:fig2app}
\end{figure*}
\begin{figure*}[tph]
	\includegraphics[width=0.42\textwidth]{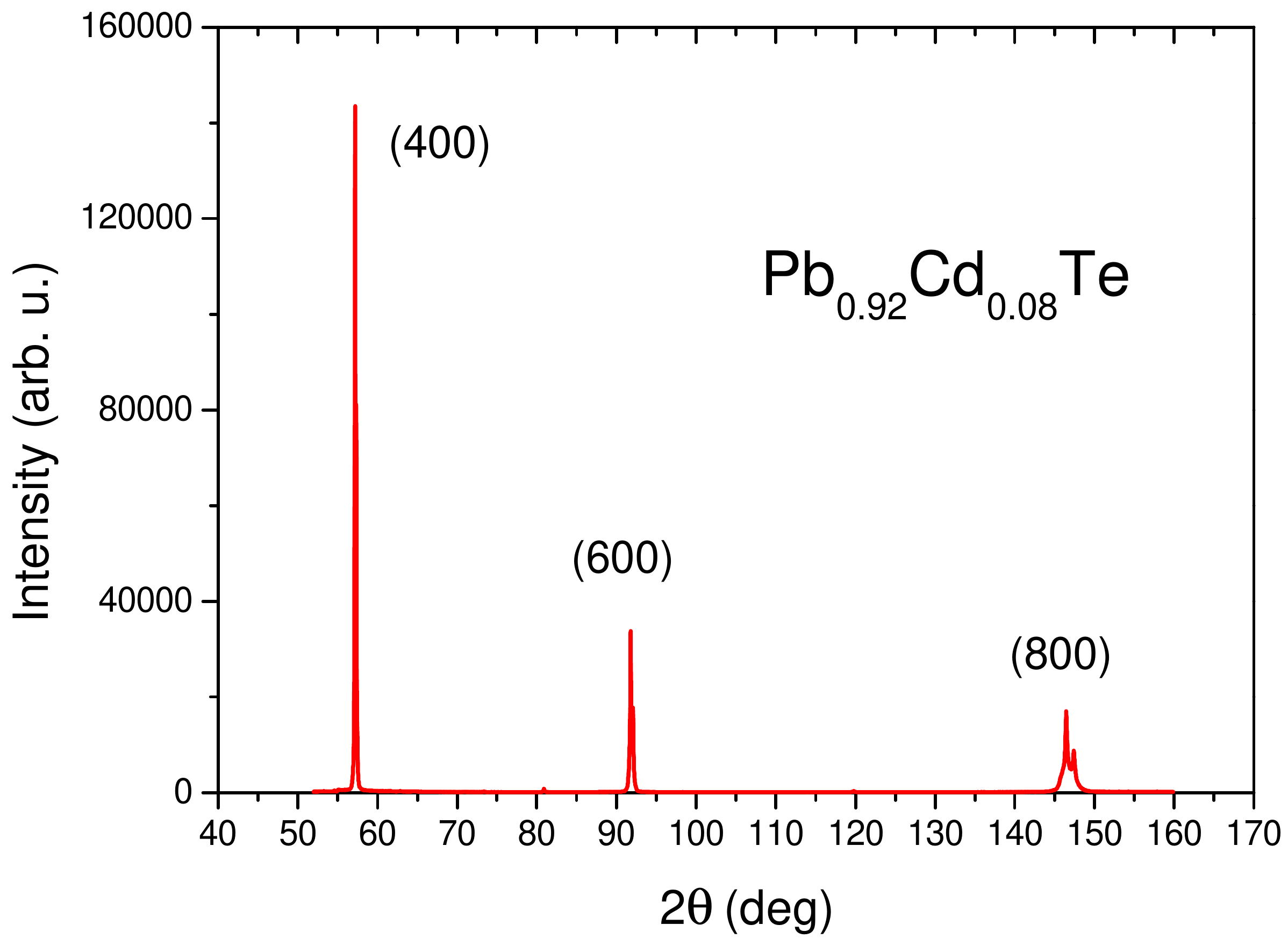}
	\caption{X-ray diffraction spectra of Pb$_{0.98}$Cd$_{0.08}$Te monocrystals.}
	\label{fig:fig2app}
\end{figure*}

\vspace*{-1em}

\section{appendix}
\begin{figure*}[tph]
	\centering
	\includegraphics[width=0.84\textwidth]{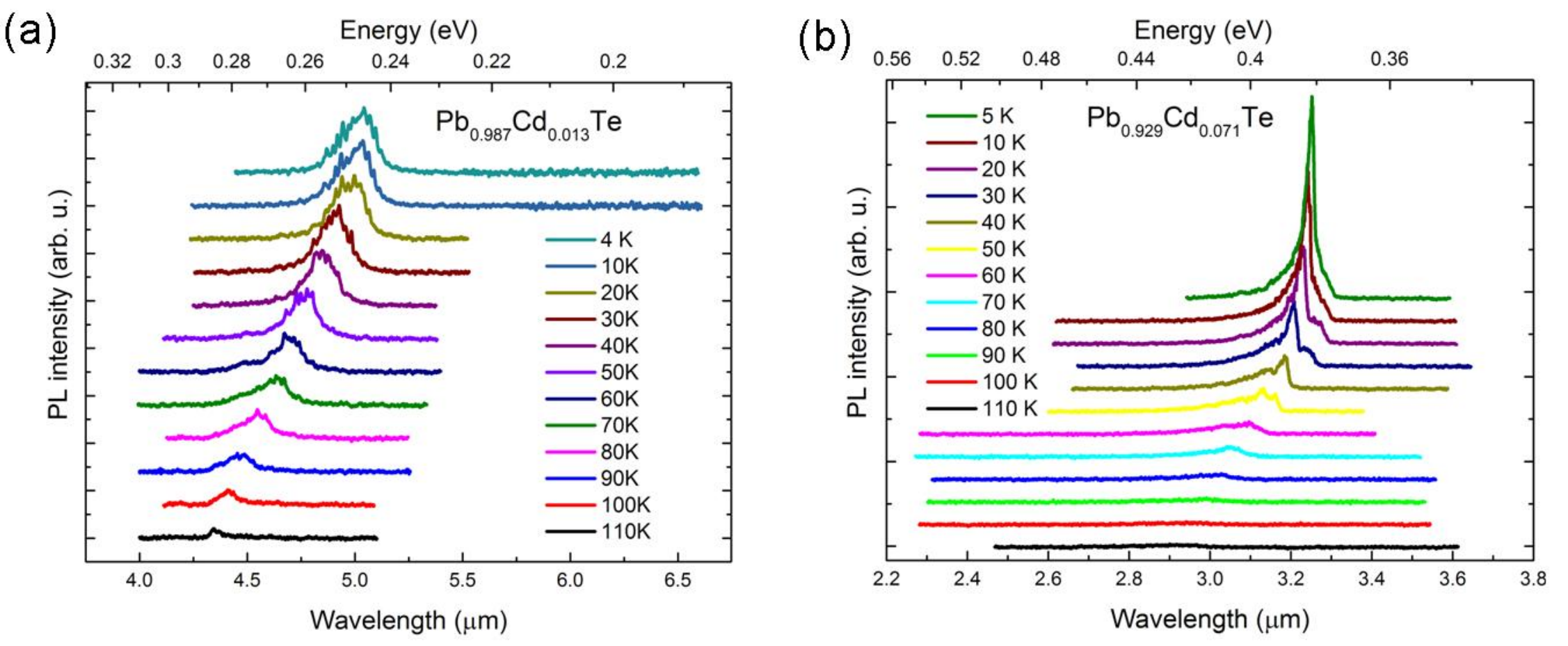}
	\caption{Photoluminesce of Pb$_{1-x}$Cd$_x$Te monocrystals with (a) $x$ = 0.013 and (b) $x$ = 0.071  measured at temperatures from 4 to 110 K.}
	\label{fig:fig2app}
\end{figure*}
\begin{figure*}[tph]
	\centering
	\includegraphics[width=0.45\textwidth]{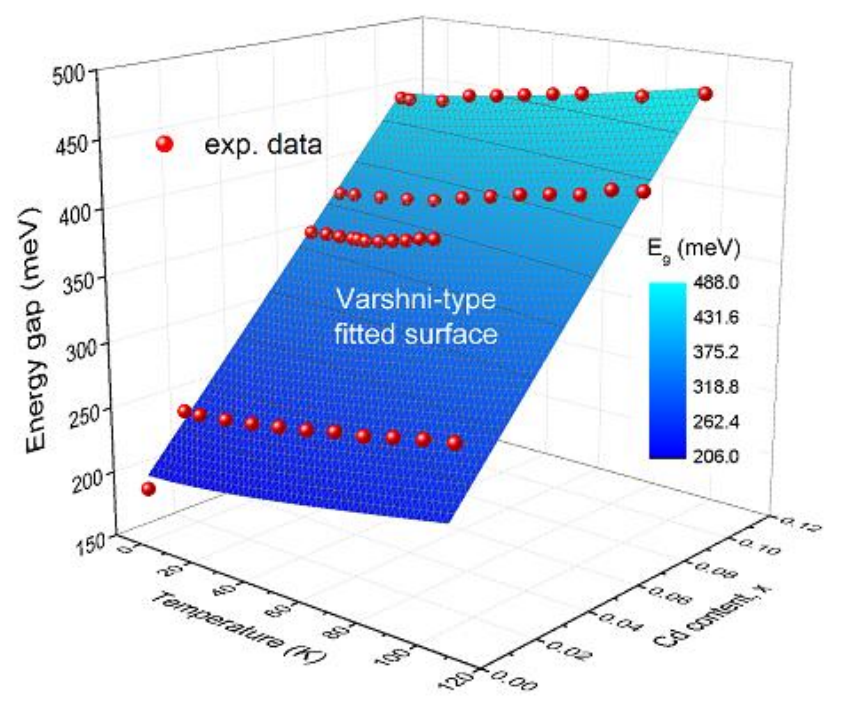}
	\caption{Dependence of E$_g$ for Pb$_{1-x}$Cd$_x$Te samples on temperature and Cd content $x$ described using Varshni-type formula (see Eq. (4)).}
	\label{fig:fig1app}
\end{figure*}
We quote below our parameters adjusted to fit all transport phenomena as well as formulas used to describe our photoluminesccence data shown in Fig. 12. For comparison we also quote $\Delta E$ and m$^*_{hh}$ values determined by other authors. Our listing does not pretend to be complete.

For PbTe at 300 K our values are:
$\Delta E$ = $E(L^+_6)-E(\Sigma_{hh})$  = 120 meV; m$^*_{hh}$ = 0.6 m$_0$; deformation potentials $E^v_{ac}=E^c_{ac}$ = 12.75 eV, $E^v_{npo}=E^c_{npo}$ = 16 eV
alloy disorder matrix elements: $E^v_{ad}$ = -2$\cdot 10^{-22}$ eVcm$^{-3}$ and $E^c_{ad}$ = 10$^{-22}$ eVcm$^{-3}$.\\

Other authors:

$\Delta E$ = $E(L^+_6)-E(\Sigma_{hh})$:

24 meV (Ref.~\onlinecite{pp72})

40 meV (Ref.~\onlinecite{pp73})

50 meV (Ref.~\onlinecite{pp74})

60 meV (Ref.~\onlinecite{pp75})

103 meV (Ref.~\onlinecite{pp76,pp79})

120 meV (Ref.~\onlinecite{pp77})

(100$\div$150) meV (Ref.~\onlinecite{pp78})\\

m$^*_{hh}$:

0.368 m$_0$ (Ref.~\onlinecite{pp83})

0.4 m$_0$ (Ref.~\onlinecite{pp80})

1 m$_0$ (Ref.~\onlinecite{pp31})

(0.6 - 2.5) m$_0$ (Ref.~\onlinecite{pp81})

\end{document}